\documentclass[aps,prd,twocolumn,amsmath,amssymb,superscriptaddress]{revtex4-2}
\usepackage{graphicx} % Required for inserting images
\usepackage{graphicx,color,dcolumn,booktabs,bm}
\usepackage{indentfirst}
\usepackage{epsfig}
\usepackage{feynmf}   %{feynmp}
\usepackage{epstopdf}   %{feynmp}
\usepackage{slashed}  %for Feynman symbols
\usepackage{enumitem}
\usepackage{cases}
\usepackage{color}
\usepackage{multirow}
\usepackage[colorlinks, citecolor=blue,menucolor=red, linkcolor=red]{hyperref}
\usepackage{verbatim}
\usepackage{mathrsfs}
\usepackage{rotating}
\usepackage{makecell}
\usepackage{threeparttable}
\usepackage{enumerate}
\usepackage{gensymb}
\usepackage{subfigure}
\usepackage{float}
\usepackage{setspace}
\usepackage{titlesec}
\usepackage{titletoc}
\usepackage{appendix}
\usepackage{hhline}

\begin{document}

%\preprint{}

\title{\boldmath Monte Carlo Simulation-Based Partial Wave Analysis of \texorpdfstring{$\Xi_c^+ \to \Xi^-\pi^+\pi^+$}{Lg} Decay}


%\thanks{}

\author{S.~X.~Li}\email{lisuxian@fudan.edu.cn}\affiliation{School of Physics, Zhengzhou University, Zhengzhou 450001, China}\affiliation{Key Laboratory of Nuclear Physics and Ion-beam Application (MOE) and Institute of Modern Physics, Fudan University, Shanghai 200443, China}
\author{R.~G.~Ping}\email{pingrg@ihep.ac.cn}\affiliation{Institute of High Energy Physics, Chinese Academy of Sciences, Beijing 100049, China}
\author{C.~P.~Shen}\email{shencp@fudan.edu.cn}\affiliation{Key Laboratory of Nuclear Physics and Ion-beam Application (MOE) and Institute of Modern Physics, Fudan University, Shanghai 200443, China}\affiliation{School of Physics, Zhengzhou University, Zhengzhou 450001, China}

\date{\today}

\begin{abstract}
  We present a Monte Carlo simulation-based partial wave analysis of the decay $\Xi_c^+ \to \Xi^-\pi^+\pi^+$ by using the Feynman-Diagram-Calculation framework. The consistency of the input and output parameters implies the reliability of the partial wave analysis method. A robust method of the spin-parity determination of the $\Xi^{*0}$ resonance is introduced. The rejection significance of the alternative spin-parity hypothesis over the favored hypothesis is strong for different spins, but weak for different parities based on the current integrated luminosity of the Belle experiment. Further data collection is needed to improve the sensitivity of parity determination in the future, e.g., Belle II experiment.
\end{abstract}

%\keywords{}

\maketitle

\section{\boldmath Introduction}
Light hadrons, composed of quarks and gluons, serve as fundamental probes into the nature of strong interactions, the underlying structure of matter, and the exploration of excited resonances. Researches on light hadrons not only provide crucial verification of Quantum Chromodynamics (QCD) in the confinement domain but also deepen our understanding of the fundamental particles and their interactions.

The ground $\Xi$ state, consisting of one light quark and two strange quarks, has been well established for several decades. However, our knowledge about its excited state $\Xi^*$ remains relatively limited. Phenomenological QCD-inspired models predict the existence of over than thirty $\Xi^*$ resonances~\cite{theo1,Wang:2024jyk}, yet only nearly ten potential candidates have been observed to date. Among these, only a few exhibit masses and widths consistent with theoretical predictions, and even fewer have well-established spin-parity ($J^P$) assignments~\cite{pdg}. According to the 2024 edition of the Review of Particle Physics (RPP)~\cite{pdg}, only the $\Xi(1530)$ has been firmly established experimentally, holding a 4-star status. Four additional states, $\Xi(1690)$, $\Xi(1820)$, $\Xi(1950)$, and $\Xi(2030)$, are classified as 3-star resonances.  Three candidates, $\Xi(1620)$, $\Xi(2250)$, and $\Xi(2370)$, hold 2-star ratings, while two states $\Xi(2120)$ and $\Xi(2500)$ remain as 1-star entries, indicating limited experimental evidence.   

Early insights into $\Xi^*$ resonances were obtained exclusively from bubble chamber experiments with very limited statistics~\cite{first-xi1620-1,first-xi1620-2,first-xi1620-3,first-1690,confirm-1690,first-xi1820,confirm-xi1820-1,confirm-xi1820-2,best-xi1950,first-xi2030,confirm-xi2030}. It wasn't until the 1980s that electron-induced experiments began to yield sufficient data. Recent years, charmed baryon decays have emerged as a very powerful tool in studying $\Xi^*$ hyperons~\cite{Li:2025exm,Li:2023olv,Liu:2023jwo,Miyahara:2016yyh}. The $\Xi(1530)$ is a well-established $\Xi^*$ state that has been clearly observed in various production mechanisms, including $K^-$-, $\Xi^-$-, and $\gamma$-induced reactions, as well as in the decays of charmed baryons~\cite{xi1530-gam,observ-xi1530,confirm-xi1620}, with the $J^P=\frac{3}{2}^+$ determined~\cite{observ-xi1530, jp-xi1530-1,jp-xi1530-2}. Evidence for the $\Xi(1690)^0$ was first reported by the Belle Collaboration in 2002 through the $\Lambda_c^+ \to \Xi(1690)^0(\to\Sigma^+K^-) K^+$ decay~\cite{evidence-1690bel}, and later confirmed by the FOCUS and BaBar Collaborations in the $\Lambda_c^+ \to \Xi(1690)^0(\to\Lambda \bar{K}^0) K^+$ decay~\cite{evidence-1690foc,obser-1690bar}. In 2019, the Belle Collaboration reported the first observation of the $\Xi(1620)^0$ and the $4.0\sigma$ evidence for the $\Xi(1690)^0$ in the $\Xi_c^+ \to \Xi^-\pi^+\pi^+$ decay~\cite{confirm-xi1620}.  In 2023, the GlueX Collaboration reported the observation of the $\Xi(1820)$ in the photoproduction process~\cite{xi1820glue}. Recently, the BESIII Collaboration observed signals for the $\Xi(1690)$ and $\Xi(1820)$ in the reaction $\psi(3686) \to K^-\Lambda\bar{\Xi}^+$ and determined the $J^P =\frac{1}{2}^-$ and $\frac{3}{2}^-$ for the $\Xi(1690)$ and $\Xi(1820)$, respectively~\cite{xi16901820-jp}. The $J^P$ quantum number of the $\Xi(1620)$ remains undetermined to date. The $\Xi(1950)$ has only been convincingly observed in hyperon-induced interactions~\cite{confirm-1690,best-xi1950}, with the unmeasured $J^P$ value. The weak evidence for the $\Xi(2030)$ was observed in the $K^-p \to K^+ X^-$~\cite{confirm-xi2030}, but no further evidence has been reported since then. A moment study of the decay angular distribution of the $\Sigma^-\bar{K}^0$ system suggested the $J \geq \frac{5}{2}$ for the $\Xi(2030)$~\cite{first-xi2030}. The existence of the remaining four $\Xi^*$ states with masses above 2100 MeV/$c^2$ is either questionable or urgently requires independent confirmation. Moreover, their $J^P$ quantum numbers are entirely unknown. 

  While phenomenological models can successfully reproduce the properties of the well-established $\Xi(1530)$, significant discrepancies arise in their predictions for higher excitations. Theoretical predictions for the masses vary by 100 to 200 MeV/$c^2$ across different frameworks, highlighting the persistent challenges in modeling strange quark dynamics within non-perturbative QCD. The $\Xi(1620)$ is strongly coupled to $\Xi \pi$, and has a large width compared to other known $\Xi^*$ states, whereas the $\Xi(1690)$ is strongly coupled to $\Sigma\bar{K}/\Lambda\bar{K}$ and has a presumably narrow width of $\Gamma \approx 10$ MeV. The large width of the $\Xi(1620)$ may indicate the presence of more than one state or suggest a more complex and exotic interpretation of the $\Xi(1620)$. The $J^P=\frac{1}{2}^-$ assignment for the $\Xi(1620)$ is supported by calculations in the Skyrme model~\cite{ref272} and chiral unitary approaches~\cite{ref267,ref268,ref274}. The classification of the $\Xi(1690)$ and $\Xi(1820)$ as $J^P=\frac{1}{2}^-$ and $J^P=\frac{3}{2}^-$ states, respectively, is strongly supported by a majority of theoretical approaches~\cite{ref269,ref15,ref270}. Unlike the other relatively narrow resonances, the $\Xi(1950)$ exhibits a significantly broader width, indicating a more complex underlying structure. The authors of Ref.~\cite{ref279} proposed searching for the $J^P=\frac{5}{2}^-$ state as a broader structure in the $\Xi\pi$ spectrum, or the $J^P=\frac{5}{2}^+$ state as a narrow structure in the $\Lambda K$ spectrum. The $\Xi(2030)$ is predominantly interpreted as a $J^P=\frac{5}{2}^-$ state, and is likely part of an octet with partners $N(1675)$, $\Lambda(1830)$, and $\Sigma(1775)$~\cite{ref279}.

  A golden channel to study properties of the $\Xi^{*0}$ states is the $\Xi_c^+ \to \Xi^-\pi^+\pi^+$. The Belle Collaboration conducted a study of this process in 2019 using a data sample with an integrated luminosity of 980 $\text{fb}^{-1}$~\cite{confirm-xi1620}. The $\Xi(1620)^0$, $\Xi(1690)^0$, and $\Xi(1530)^0$ were seen in the $M(\Xi^-\pi^+)$ spectrum, as shown in Fig.~1 of Ref.~\cite{confirm-xi1620}. An unknown structure in the range 1.8$-$2.1 GeV/$c^2$ was also seen, which was attributed to resonances such as $\Xi(1820)^0$, $\Xi(1950)^0$, and $\Xi(2030)^0$. The reflection of the $\Xi(1530)^0 \to \Xi^-\pi^+_L$ decay produced a peak around 2.2 GeV/$c^2$ in the $M(\Xi^-\pi^+_H)$ spectrum,  where the pion with the lower (higher) momentum is labeled $\pi^+_L~(\pi^+_H)$. A significant nonresonant contribution was also taken into account in the fit to the $M(\Xi^-\pi^+_L)$ distribution displayed in Fig.~2 of Ref.~\cite{confirm-xi1620}. Partial wave analysis (PWA) is one of the most powerful techniques for studying the internal dynamics of three-body decays. It enables researchers to decompose the decay process into different angular momentum components, offering insights into the underlying physics, including resonant contributions, spin-parity properties, and interactions between the final-state particles. For the decay $\Xi_c^+ \to \Xi^-\pi^+\pi^+$, the large statistics collected by the Belle experiment make it particularly suitable for PWA.

  In this work, we demonstrate the reliability of PWA based on the Feynman-Diagram-Calculation (FDC) framework~\cite{fdcpwa} for analyzing the $\Xi_c^+ \to \Xi^-\pi^+\pi^+$ decay using the toy Monte Carlo (MC) simulations and introduce a robust method for determining the $J^P$ value of the $\Xi^{*0}$ resonance. FDC automatically constructs effective Lagrangian, deduces Feynman rules for the Standard Model, handles the phase space (PHSP) integration for processes with multiple final-state particles, and calculates the squared amplitude for a decay. 

\section{Decay amplitude}
  The decay amplitude for the process $\Xi_c^+ \to \Xi^{*0}\pi^+$ with $\Xi^{*0} \to \Xi^-\pi^+$ is constructed using the covariant tensor formalism~\cite{covariant-tensor}, expressed as 
  \begin{equation}
  \begin{aligned}
    M_i=&\bar{u}(\vec{p}_{\Xi^-},\lambda_{\Xi^-})V_{\Xi^{*0}\to\Xi^-\pi^+}G^{(J^P)}\times\\
    &V_{\Xi_c^+\to\Xi^{*0}\pi^+}u(\vec{p}_{\Xi_c^+},\lambda_{\Xi_c^+}),
  \end{aligned}
  \end{equation}
  where $u(\vec{p},\lambda)$ is the spinor for a baryon, $\vec{p}$ and $\lambda$ are the momentum and the third component of the spin projection, respectively, $G^{(J^P)}$ is the propagator of the $\Xi^{*0}$ resonance, and $V$ is the effective vertex. The total decay amplitude of the $\Xi_c^+ \to \Xi^-\pi^+\pi^+$ process is calculated as $\mathcal{M}=\sum_ic_iM_i$, where $M_i$ is the amplitude for the $i$-th resonance, and $c_i$ is the complex parameter and includes the information of complex coupling constants $f_i$, as introduced below.
  
  The spinor for a spin-1/2 baryon is expressed as:
    \begin{equation}
    u(\vec{p},\lambda)=\begin{pmatrix}\sqrt{E+m}\\\frac{\vec{\sigma}\cdot\vec{p}}{\sqrt{E+m}}\end{pmatrix}\chi(\lambda),
    \end{equation}
    where $m$, $E$, and $\vec{p}$ are the mass, energy, and momentum of the baryon, respectively, $\vec{\sigma}$ is the pauli vector, and $\chi(\lambda)$ is the spin wave-function for a baryon. 
    The projection operator for a spin-1/2 baryon is defined as:
    \begin{equation}
    \mathcal{P}^{(\frac{1}{2})}=\sum_{\lambda}u(\vec{p},\lambda)\bar{u}(\vec{p},\lambda)=\slashed{p}+m,
    \end{equation}
    where $\slashed{p}\equiv\gamma^{\mu}p_{\mu}$. The spinor for a spin-($n+\frac{1}{2}$) baryon ($n=1,~2,~3,~...$) can be obtained using the polarization vector and the Clebsch-Gordan coefficient~\cite{spinorH0,spinorH1,spinorH2}, as follows:
    \begin{widetext}
    \begin{equation}
    \begin{aligned}
    U_{\mu_1\mu_2...\mu_n}(\vec{p},\lambda)&=\sum_{\lambda_n,\lambda_{n+1}}\left<n,\lambda_n;\frac{1}{2},\lambda_{n+1}|n+\frac{1}{2},\lambda\right>\epsilon_{\mu_1\mu_2...\mu_n}(\vec{p},\lambda_n)u(\vec{p},\lambda_{n+1}),\\
    \epsilon_{\mu_1\mu_2...\mu_n}(\vec{p},\lambda)&=\sum_{\lambda_{n-1},\lambda_n}\left<n-1,\lambda_{n-1};1,\lambda_n|n,\lambda\right>\epsilon_{\mu_1\mu_2...\mu_{n-1}}(\vec{p},\lambda_{n-1})\epsilon_{\mu_n}(\vec{p},\lambda_n),
    \end{aligned}   
    \end{equation}
    \end{widetext}
    where $\epsilon_{\mu_n}(\vec{p},\lambda_n)$ is the polarization vector. 
    The wave function satisfies the Rarita-Schwinger conditions~\cite{rasch-condi}.
    The projection operator for a spin-($n+\frac{1}{2}$) baryon ($n=1,~2,~3,~...$)~\cite{spinorH1,spinorH2} is given by:
    \begin{small}
    \begin{equation}
    \begin{aligned}
    \mathcal{P}^{(n+\frac{1}{2})}_{\mu_1\mu_2...\mu_n,\nu_1\nu_2...\nu_n}&=\sum_{\lambda}U_{\mu_1\mu_2...\mu_n}(\vec{p},\lambda)\bar{U}_{\nu_1\nu_2...\nu_n}(\vec{p},\lambda)\\
      &=\frac{n+1}{2n+3}(\slashed{p}+m)\gamma^{\alpha}\gamma^{\beta}\mathcal{P}^{(n+1)}_{\alpha\mu_1\mu_2...\mu_n,\beta\nu_1\nu_2...\nu_n}, 
    \end{aligned}
    \end{equation}
    \end{small}
    where
    \begin{small}
    \begin{equation}
    \mathcal{P}^{(n)}_{\mu_1\mu_2...\mu_n,\nu_1\nu_2...\nu_n}=\sum_{\lambda}\epsilon_{\mu_1\mu_2...\mu_n}(\vec{p},\lambda)\epsilon^{*}_{\nu_1\nu_2...\nu_n}(\vec{p},\lambda).
    \end{equation}
    \end{small}

  The $G^{(J^P)}$ is constructed using the projection operator and the relativistic Breit-Wigner (BW) function with a mass-dependent width, expressed as:
  \begin{equation}
  G^{(n+\frac{1}{2})}=\frac{2M_0}{M^2-M_0^2+iM_0\Gamma(M)}F_b\mathcal{P}^{(n+\frac{1}{2})},
  \end{equation}
  where $n=0,~1,~2,~...$, $M$ is the invariant mass of $\Xi^-\pi^+$, and $M_0$ and $\Gamma(M)$ are the mass and width of the $\Xi^{*0}$, respectively. Since baryons are not point-like particles, form factors modifying the BW shape are required to describe them. We use a phenomenological form factor~\cite{formfac1,formfac2} for the resonance, defined as 
  \begin{equation}
  F_b=
  \left\{
  \begin{aligned}
  &1 & J=1/2 \\
  &e^{-\frac{|M^2-M_0^2|}{8M_0\Gamma}} & J=3/2 \\
  &e^{-\frac{|M^2-M_0^2|}{4M_0\Gamma}} & J=5/2 \\
  &e^{-\frac{|M^2-M_0^2|}{2M_0\Gamma}} & J=7/2
  \end{aligned}
  \right. .
  \end{equation}
Different form factors, such as those used in Ref.~\cite{xi16901820-jp}, are also tested and the choices of the different form factor have no significant impact on the PWA results.

The effective vertices are deduced from an effective Lagrangian $\mathcal{L}=\bar{\psi_1}V\psi_2A$, where $\psi_{1,2}$ and $A$ represent the baryon and meson fields, respectively.  For a strong interaction, the corresponding interaction Lagrangian must be Lorentz invariant, $C$-parity invariant, $P$-parity invariant,  and $CPT$ invariant. These constraints imply that the strong interaction vertices must satisfy
  \begin{equation}
  \begin{aligned}
    V&=\zeta_A C(\gamma_0V^{\dagger}\gamma_0)^TC^{-1},\\
    V&=\eta^*_1\eta_2\eta_A\gamma_0V^P\gamma_0,
  \end{aligned}
  \end{equation}
  where $\zeta_A$ is the $C$-parity of the meson, $\eta_A$ and $\eta_{1,2}$  are the $P$-parities of the meson and baryon, respectively, and $C=-i\gamma_2\gamma_0$ is the charge conjugate transform operator. 
  We enumerate several types of vertices involved in baryon decays, such as $V=i,~\gamma_5,~\gamma_\mu,~ \gamma_\mu\gamma_5,~\sigma_{\mu\nu}={i\over 2}(\gamma_\mu\gamma_\nu-\gamma_\nu\gamma_\mu),~\sigma_{\mu\nu}\gamma_5$, and $g_{\mu\nu}$, and their transformations under charge-conjugation and $C$ parity are provided in Table~\ref{tab:trans}.
  \begin{table}[htbp]
  \centering
  \caption{The transformation properties of some operators.\label{tab:trans}}
  \begin{tabular}{cccccccc}
  \hline\hline
  $V$     & $i$ &$\gamma_5$ &$\gamma_\mu$ &$\gamma_\mu\gamma_5$ &$\sigma_{\mu\nu}$ &$\sigma_{\mu\nu} \gamma_5$ &$g_{\mu\nu}$ \\\hline
  $\gamma_0V^\dagger\gamma_0$     & $-i$&$-\gamma_5$& $\gamma_\mu$&$\gamma_\mu\gamma_5$&$\sigma_{\mu\nu}$&$- \sigma_{\mu\nu} \gamma_5$&$g_{\mu\nu}$ \\
  $C(\gamma_0 V^\dagger\gamma_0)C^{-1}$&$-i$&$-\gamma_5$&$-\gamma_\mu$&$\gamma_\mu\gamma_5$&$-\sigma_{\mu\nu}$&$\sigma_{\mu\nu} \gamma_5$&$g_{\mu\nu}$ \\
  $\gamma_0V^P\gamma_0$ & $i$&$-\gamma_5$&$\gamma_\mu$& $-\gamma_\mu\gamma_5$&$\sigma_{\mu\nu}$& $-\sigma_{\mu\nu} \gamma_5$&$g_{\mu\nu}$ \\\hline\hline
  \end{tabular}
   \end{table}
   
  The first decay, $\Xi_c^+ \to \Xi^{*0}\pi^+$, does not conserve $P$-parity due to the quark-level transformation $b\to cs$. As a result, parity conservation is not required in this decay. In contrast, the second decay $\Xi^{*0} \to \Xi^-\pi^+$ conserves $P$-parity because it proceeds via a strong interaction.
  Table~\ref{tab:vertice} lists effective vertices of $\Xi_c^+(p) - \Xi^{*0}[J^P] - \pi^+(p_1)$ and $\Xi^{*0}[J^P] - \Xi^-(p_2) - \pi^+(p_3)$ with the $J^P = \frac{1}{2}^-$, $\frac{1}{2}^+$, $\frac{3}{2}^-$, $\frac{3}{2}^+$, $\frac{5}{2}^-$, $\frac{5}{2}^+$, $\frac{7}{2}^-$, and $\frac{7}{2}^+$, where $p$ and $p_{1,2,3}$ denote the four-momenta of the initial and final state particles, respectively. Resonances with spin $J\geq\frac{9}{2}$ are not considered in this work due to suppression effects.  

  \begin{table}%[H]
     \centering
     \caption{Effective vertices $\Xi_c^+(p) - \Xi^{*0}[J^P] - \pi^+(p_1)$ and $\Xi^{*0}[J^P] - \Xi^-(p_2) - \pi^+(p_3)$ with the $J^P = \frac{1}{2}^-$, $\frac{1}{2}^+$, $\frac{3}{2}^-$, $\frac{3}{2}^+$, $\frac{5}{2}^-$, $\frac{5}{2}^+$, $\frac{7}{2}^-$, and $\frac{7}{2}^+$. The $f_{i~(i=1,~2,~3,~...)}$ is the complex parameter, usually determined by the fit to data. The $\mu$, $\nu$, $\alpha$, and $\beta$ are the Lorentz indexes. \label{tab:vertice} }
     \begin{tabular}{ccc } \hline \hline 
      $J^P$ & $\Xi_c^+(p) - \Xi^{*0} - \pi^+(p_1)$ & $\Xi^{*0} - \Xi^-(p_2) - \pi^+(p_3)$  \\ \hline
      $\frac{1}{2}^-$    &$f_{1}\gamma_5i+f_{2}$ &$f_{3}$ \\
      $\frac{1}{2}^+$    &$f_{4}\gamma_5i+f_{5}$ &$f_{6}\gamma_5i$ \\
      $\frac{3}{2}^-$    &$f_{7}p_\mu i+f_{8}\gamma_5p_\mu $ &$f_{9}\gamma_5{p_2}_\alpha$ \\
      $\frac{3}{2}^+$    &$f_{10}p_\mu i+f_{11}\gamma_5p_\mu $ &$f_{12}{p_2}_\alpha i$ \\
      $\frac{5}{2}^-$    &$f_{13}p_\mu p_\nu +f_{14}\gamma_5p_\mu p_\nu i$ &$f_{15}{p_2}_\alpha {p_2}_\beta$ \\
      $\frac{5}{2}^+$    &$f_{16}p_\mu p_\nu + f_{17}\gamma_5p_\mu p_\nu i$ &$f_{18}\gamma_5{p_2}_\alpha {p_2}_\beta i$ \\
      $\frac{7}{2}^-$    &$f_{19}p_\mu p_\nu p_\sigma i + f_{20}\gamma_5p_\mu p_\nu p_\sigma$ & $f_{21}\gamma_5{p_2}_\alpha {p_2}_\beta {p_2}_\tau$ \\
      $\frac{7}{2}^+$    &$f_{22}p_\mu p_\nu p_\sigma i + f_{23}\gamma_5p_\mu p_\nu p_\sigma$ & $f_{24}{p_2}_\alpha {p_2}_\beta {p_2}_\tau i$ \\ 
      \hline \hline
     \end{tabular}
  \end{table}
  
\section{Toy MC simulation}
 The FDC package constructs the effective Lagrangian and deduces Feynman rules from a simple input based on fundamental requirements such as isospin invariance, Lorentz invariance, strange number conservation, charm number conservation, $C$-parity invariance, $P$-parity invariance, and $G$-parity invariance. The input consists of a list of mesons and baryons along with their properties, such as $J^P$, mass, and width. The FDC then constructs the decay amplitude used to perform a likelihood fit to data. It has been successfully applied in some partial wave analyses in BESIII experiment~\cite{fdc-bes3-1,fdc-bes3-2,fdc-bes3-3,fdc-bes3-4,fdc-bes3-5}.
\subsection{Toy MC generation}
\label{subsec:toymcgene}
  We construct the effective Lagrangian of the $\Xi_c^+ \to \Xi^+ \pi^-\pi^-$ decay using FDC. The input properties of the possible $\Xi^{*0}$ states are listed in Table~\ref{tab:input-property}, where the masses, widths, and $J^P$ values of the $\Xi(1530/1820)^0$ are taken from world-averaged values~\cite{pdg}. The $J^P$ quantum number of the $\Xi(1690)^0$ is taken as $\frac{1}{2}^-$, measured by BESIII recently~\cite{xi16901820-jp}. We assume that the decay $\Xi^{*0} \to \Xi^- \pi^+$ proceeds via an $S$-wave for the $\Xi(1620/1950)^0$, as their $J^P$ values are unknown. The parity of $\Xi(2030)^0$ has been determined to be $P=-1$ with its spin satisfying $J\geq\frac{5}{2}$~\cite{pdg}. Therefore, we assume $J^P=\frac{5}{2}^-$ for the $\Xi(2030)^0$. We also tested alternative $J^P$ assignments beyond those specified in Table~\ref{tab:input-property}, including $\frac{3}{2}^-$ and $\frac{5}{2}^-$ for the $\Xi(1620/1950)^0$, as well as $\frac{7}{2}^+$ for the $\Xi(2030)^0$. Our analysis demonstrates that the choice of input $J^P$ values does not significantly impact the final conclusions. The nonresonant process of the $\Xi_c^+ \to \Xi^+ \pi^-\pi^-$ is also considered via the PHSP model. 
  \begin{table}[htbp]
     \centering
     \caption{The input properties of the $\Xi^{*0}$ states. \label{tab:input-property}}
     \begin{tabular}{c c c c } \hline \hline
      Resonance   & Mass (MeV/$c^2$) & Width (MeV) & $J^P$ \\ \hline
      $\Xi(1530)^0$ & $1531.8$ & $9.1$  & $\frac{3}{2}^+$  \\
      $\Xi(1620)^0$ & $1620.0$ & $32.0$ & $\frac{1}{2}^-$  \\
      $\Xi(1690)^0$ & $1690.0$ & $20.0$ & $\frac{1}{2}^-$ \\
      $\Xi(1820)^0$ & $1823.0$ & $24.0$ & $\frac{3}{2}^-$ \\
      $\Xi(1950)^0$ & $1950.0$ & $60.0$ & $\frac{1}{2}^-$ \\
      $\Xi(2030)^0$ & $2025.0$ & $20.0$ & $\frac{5}{2}^-$ \\ 
      \hline \hline
     \end{tabular}
  \end{table}
  Table~\ref{tab:input-ratio} summarizes the assumed ratios for each resonance and the corresponding interference ratios between pairs of resonances. These ratios are determined based on the $M(\Xi^-\pi^+)$ spectrum shown in Fig.~1 of Ref.~\cite{confirm-xi1620}.
  \begin{table*}[htbp]
     \centering
     \caption{Assumed ratio of each resonance and the corresponding interference term between two resonances (in \%). \label{tab:input-ratio}}
     \begin{tabular}{c c c c c c c c} \hline \hline
      Ratio  & $\Xi(1530)^0$ & $\Xi(1620)^0$ & $\Xi(1690)^0$ & $\Xi(1820)^0$ & $\Xi(1950)^0$ & $\Xi(2030)^0$ & nonresonance \\ \hline
      $\Xi(1530)^0$ & 5.4 & 0.0 & 0.0 & 0.0 & 0.0 & 0.1 & $-0.2$ \\
      $\Xi(1620)^0$ & & 1.3 & 0.1 & 0.0 & $-0.1$ & 0.0 & 1.4 \\
      $\Xi(1690)^0$ & & & 0.7 & 0.0 & 0.0 & 0.0 & 1.1 \\
      $\Xi(1820)^0$ & & & & 2.8 & $-0.6$ & 0.0 & 0.1 \\
      $\Xi(1950)^0$ & & & & & 9.0 & 0.0 & $-0.8$ \\
      $\Xi(2030)^0$ & & & & & & 1.5 & 0.0 \\
      Nonresonance & & & & & & & 79.3 \\
      \hline \hline
     \end{tabular}
  \end{table*}

  Using the above properties as the input parameters, we generate toy MC samples according to the amplitude through the sampling method. Approximately 45,000 $\Xi_c^+ \to \Xi^- \pi^+ \pi^+$ events with a background contamination of around 20\% are selected from a $980~\text{fb}^{-1}$ data sample collected by the Belle Collaboration~\cite{confirm-xi1620}. To ensure the reliability of the PWA results, it is preferable to suppress the background level to less than 10\%. Assuming the signal level is also reduced by half, approximately 18,000 signal events are generated. The 1,800 toy background events are generated based on the background shape reported by the Belle Collaboration~\cite{confirm-xi1620}. We also generate a large toy PHSP sample with 300,000 events to ensure the accuracy of the PWA results.  

\subsection{Fit to toy MC samples}  
  The complex coupling constants $c_i$ are determined using an unbinned maximum likelihood fit. The joint probability density for observing $N$ events in a data set is given by
  \begin{equation}
     \mathcal{L}=\prod_{i=1}^{N} P(x_i),
  \end{equation}
  where $P(x_i)$ is a probability to produce event $i$ with four-vector momentum $x_i = (E,~ p_x,~p_y,~p_z)_i$. The $P(x_i)$ is calculated from the differential cross section as follows
  \begin{equation}
     P(x_i)=\frac{\omega(x_i)}{\int \omega(x)d\Phi},
  \end{equation}
  where $\omega(x_i) = (d\sigma/d\Phi)_i$ is the differential cross section and $\int \omega(x)d\Phi$ is the observed total cross section. The differential cross section is given by
  \begin{equation}
     \omega(x) \equiv \frac{d\sigma}{d\Phi}=\left|\sum_j c_j M_j \right|^2.
  \end{equation}
  For technical reasons, rather than maximizing $\mathcal{L}$, the object function, $\mathcal{S} = -\text{ln}\mathcal{L}$, is minimized using the MINUIT package~\cite{minut}, where 
  \begin{equation}
     \mathcal{S}=-\sum_i^N\text{ln}\omega(x_i)+N\text{ln}\left[\int \omega(x)d\Phi \right]. 
  \end{equation}
  The $\int \omega(x)d\Phi$ is evaluated using a PHSP MC sample consisting of $N_\text{PHSP}$ events.  The normalization integral is then computed as follows 
  \begin{equation}
     \int \omega(x)d\Phi=\frac{1}{N_\text{PHSP}}\sum_k^{N_\text{PHSP}}\omega(x_k).
  \end{equation}
  The background events are subtracted from the $-\text{ln}\mathcal{L}$ function using $-\text{ln}\mathcal{L}=-(\text{ln}\mathcal{L}_\text{data}-\text{ln}\mathcal{L}_\text{bkg})$.

\begin{figure}[hbtp]
    \begin{center}
    \includegraphics[width=0.48\textwidth]{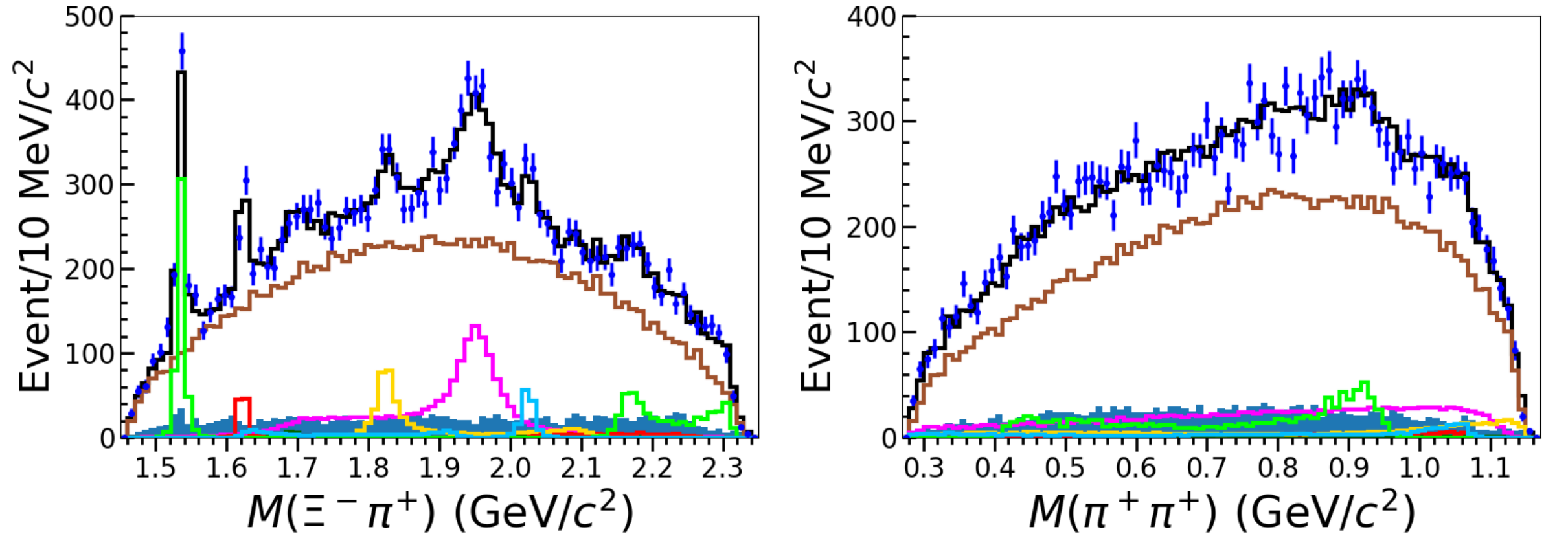}
    \includegraphics[width=0.48\textwidth]{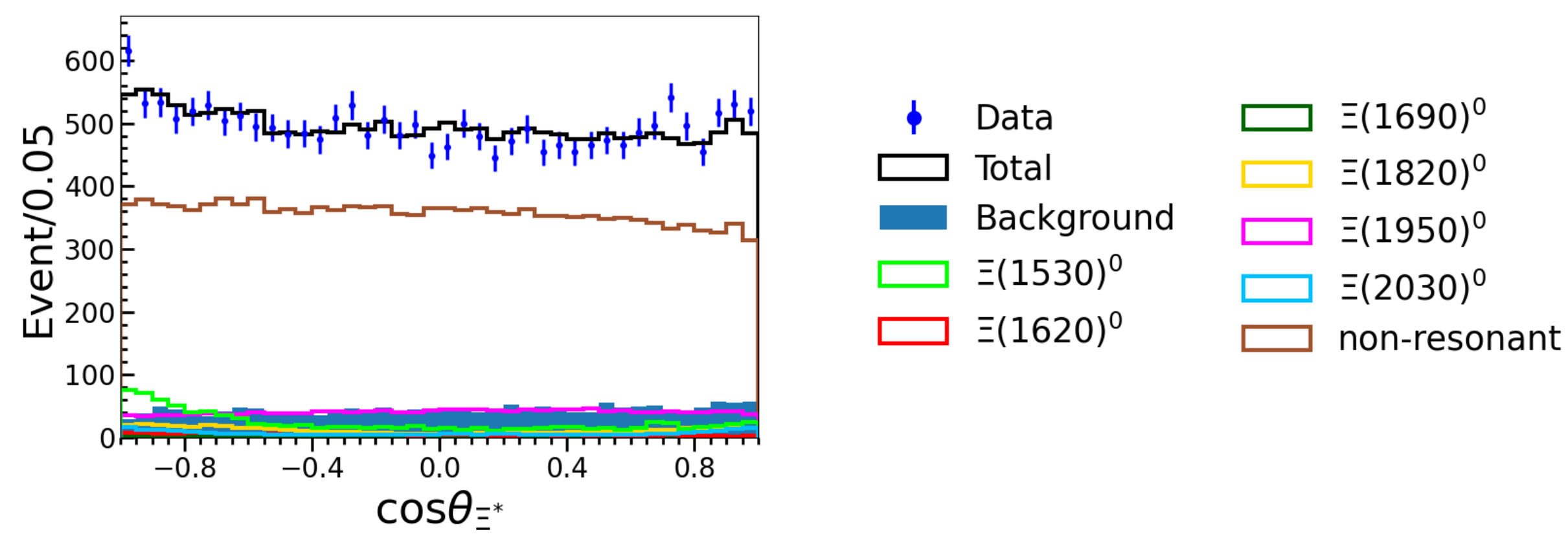}
    \caption{The $M(\Xi^-\pi^+)$, $M(\pi^+\pi^+)$, and $\cos\theta_{\Xi^{*0}}$ distributions from the best fit result. The dots with error bars are the toy MC samples; the black curves are from the best fit result; the blue shaded areas represent the background events; the lime, red, dark-green, gold, magenta, deep-sky-blue, and sienna curves represent the $\Xi(1530)^{0}$, $\Xi(1620)^{0}$, $\Xi(1690)^{0}$, $\Xi(1820)^{0}$, $\Xi(1950)^{0}$, $\Xi(2030)^{0}$, and nonresonance, respectively.}
    
    \label{fig:sec4:mxipi-best}
    \end{center}
  \end{figure}

  We perform an unbinned maximum likelihood fit to the toy MC samples to verify the consistence between the input parameters and the fitted output values. Figure~\ref{fig:sec4:mxipi-best} shows the distributions of $M(\Xi^-\pi^+)$, $M(\pi^+\pi^+)$, and $\cos\theta_{\Xi^{*0}}$ from the best fit result, where $\theta_{\Xi^{*0}}$ is the angle between the $\Xi^-$ momentum vector in the $\Xi^{*0}$ rest frame and the opposite of the $\Xi_c^+$ momentum vector in the laboratory frame. The reduced $\chi^2$ values of the fit are $\chi^2/\rm d.o.f=93.9/87=1.08$ for $M(\Xi^-\pi^+)$, $114.3/87=1.31$ for $M(\pi^+\pi^-)$, and $46.5/40=1.16$ for $\cos\theta_{\Xi^{*0}}$, where $\rm d.o.f$ denotes the number of degrees of freedom. 

\begin{figure}[htbp]
    \begin{center}
    \includegraphics[width=0.40\textwidth]{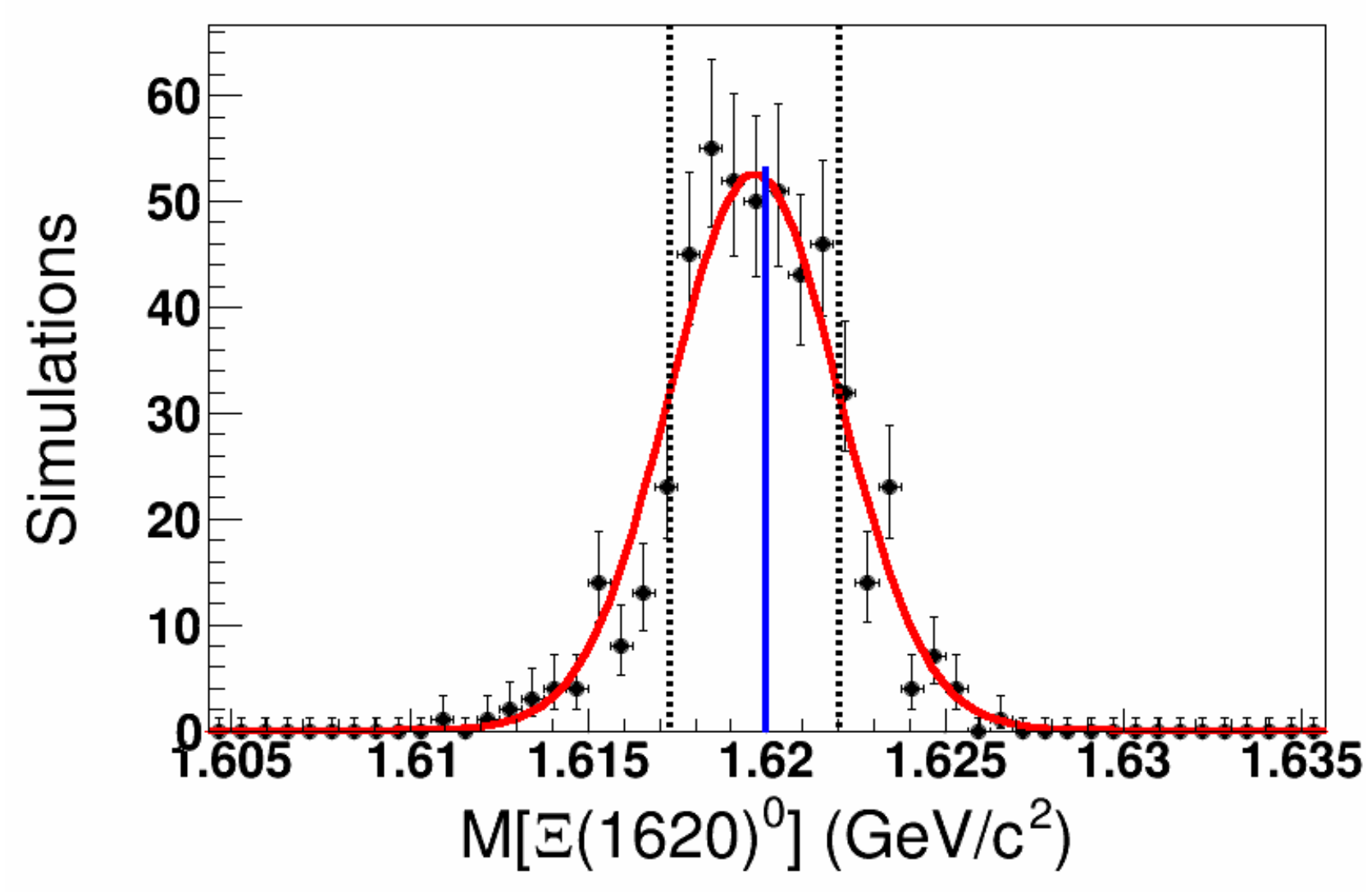}
    \caption{The distribution of the fitted masses of the $\Xi(1620)^0$. The dots with error bars are the fitted values, the red curve is the Gaussian function, the two black dashed lines correspond to $\pm1$ standard deviation, and the blue line represents the input value. }
    \label{fig:sec5:ioc}
    \end{center}
    \end{figure}

  Five hundred sets of toy samples are generated, each with the same size as described in Sec~\ref{subsec:toymcgene}. We fit these sets of toy samples and extract the ratios, masses, and widths for the input components. The distributions of these parameters for each resonance are approximately Gaussian. Figure~\ref{fig:sec5:ioc} displays the mass distribution of the $\Xi(1620)^0$ as an example, where dots with error bars represent the output values, fitted with a Gaussian function shown by the red curve, the two black dashed lines indicate $\pm1$ standard deviation derived from the Gaussian fit, and the blue line represents the input value. The fitted mean value of the Gaussian function is consistent with the input value. The distributions of other variables are provided in the Appendix.

\section{Spin-parity determination}
  The $J^P$ assignments $\frac{1}{2}^{\pm}$, $\frac{3}{2}^{\pm}$, $\frac{5}{2}^{\pm}$, and $\frac{7}{2}^{\pm}$ are considered in the fit for the resonance $\Xi^{*0}$. The minimum log-likelihood values are determined for each $J^P$ hypothesis, and the one with the smallest log-likelihood is selected as the favored $J^P$ assignment. The favored ones are consistent with the input values listed in Table~\ref{tab:input-property}. 

  A likelihood ratio $t$ is used as a test variable to discriminate between the favored $J^P$ hypothesis ($J^P_\text{fav}$) and the alternative $J^P$ hypotheses ($J^P_\text{alt}$), calculated as $t\equiv -2\text{ln}\left[\mathcal{L}(J^P_\text{alt})/\mathcal{L}(J^P_\text{fav})\right]$. The $t$ distribution is obtained from a series of toy simulated experiments, following the method described  in Ref.~\cite{ttest-method}. The toy simulated sample for each hypothesis is generated based on its joint probability density. Typically the $t$ distribution with a minus mean value is obtained under the $J^P_\text{alt}$ hypothesis, while a positive mean value corresponds to the $J^P_\text{fav}$ hypothesis. We perform the $t$-test to estimate the discrimination power of the $J^P_\text{fav}$ over $J^P_\text{alt}$.  The test statistic $t$  distributions for the $\Xi(1620)^0$ are shown in Fig.~\ref{fig:sec4:ttest-1620}, while those for others are shown in the Appendix. The simulations represented by the red dots with error bars are conducted under the $J^P_\text{fav}$ hypothesis, while those represented by the blue dots with error bars correspond to the $J^P_\text{alt}$ hypothesis. 

  \begin{figure*}[htbp]
    \begin{center}
    \includegraphics[width=0.98\textwidth]{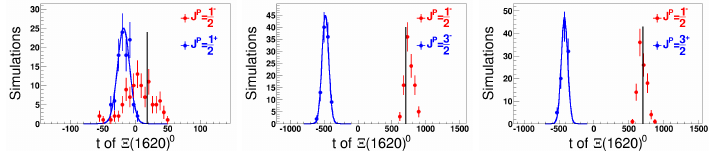}
    \caption{Distributions of the test statistic $t$ for the $\Xi(1620)^0$, for simulated experiments under the $J^P_\text{alt}=\frac{1}{2}^{+},~\frac{3}{2}^{-}$, and $\frac{3}{2}^+$ hypotheses (blue histograms) and the $J^P_\text{fav}=\frac{1}{2}^-$ hypothesis (red histograms). The values of the test statistics for the toy data, $t_\text{toydata}$, are shown by the solid vertical lines. The curves show the Gaussian fits to the distributions of the left peaks.}
    \label{fig:sec4:ttest-1620}
    \end{center}
  \end{figure*} 

  A statistical significance for rejecting the $J^P_\text{alt}$ hypothesis in favor of $J^P_\text{fav}$ hypothesis is estimated as $(t_\text{toydata}-\left<t\right>)/\sigma(t)$~\cite{ttest-method}, where $\left<t\right>$ and $\sigma(t)$ are the mean and standard deviation of the $t$ distribution under the $J^P_\text{alt}$ hypothesis, obtained by the fit with a single Gaussian function to the $t$ distribution. The $t_\text{toydata}$ is the $t$ value obtained from the toy data, which is positive and favors the $J^P_\text{fav}$ assignment. Table~\ref{tab:ttest-signif} summarizes the significances of rejection of $J^P_\text{alt}$ for each resonance. 

  \begin{table}[htbp]
     \centering
     \caption{The significances of rejection of $J^P_\text{alt}$ for each resonance. \label{tab:ttest-signif}}
     \renewcommand{\arraystretch}{1.4}
     \begin{footnotesize}
     \begin{tabular}{c c c c c c c c c c} \hline \hline
      \multirow{2}{*}{Resonance} & \multirow{2}{*}{$J^P_\text{fav}$} & \multicolumn{5}{c}{$J^P_\text{alt}$} \\ & & \normalsize{$\frac{1}{2}^-$} & \normalsize{$\frac{1}{2}^+$} & \normalsize{$\frac{3}{2}^-$} & \normalsize{$\frac{3}{2}^+$} & \normalsize{$\frac{5}{2}^-$} & \normalsize{$\frac{5}{2}^+$} & \normalsize{$\frac{7}{2}^-$} & \normalsize{$\frac{7}{2}^+$} \\ \hline
      \footnotesize{$\Xi(1530)^0$} & \normalsize{$\frac{3}{2}^+$} & \footnotesize{$10.0\sigma$} & \footnotesize{$9.0\sigma$} & \footnotesize{$2.0\sigma$} & $\cdots$ & \footnotesize{$5.8\sigma$} & \footnotesize{$6.6\sigma$} & $\cdots$ & $\cdots$ \\ 
      $\Xi(1620)^0$ & \normalsize{$\frac{1}{2}^-$} & $\cdots$ & $3.9\sigma$ & $22.4\sigma$ & $24.5\sigma$ & $\cdots$ & $\cdots$ & $\cdots$ & $\cdots$  \\ 
      $\Xi(1690)^0$ & \normalsize{$\frac{1}{2}^-$} & $\cdots$ & $1.8\sigma$ & $24.3\sigma$ & $22.1\sigma$ & $\cdots$ & $\cdots$ & $\cdots$ & $\cdots$ \\
      $\Xi(1820)^0$ & \normalsize{$\frac{3}{2}^-$} & $28.0\sigma$ & $26.5\sigma$ & $\cdots$ & $2.8\sigma$ & $5.8\sigma$ & $5.7\sigma$ & $\cdots$ & $\cdots$  \\ 
      $\Xi(1950)^0$ & \normalsize{$\frac{1}{2}^-$} & $\cdots$ & $3.2\sigma$ & $23.3\sigma$ & $23.8\sigma$ & $\cdots$ & $\cdots$ & $\cdots$ & $\cdots$ \\
      $\Xi(2030)^0$ & \normalsize{$\frac{5}{2}^-$} & $\cdots$ & $\cdots$ & $6.3\sigma$ & $6.0\sigma$ & $\cdots$ & $2.9\sigma$ & $11.6\sigma$ & $8.7\sigma$ \\ 
      \hline \hline
     \end{tabular}
     \end{footnotesize}
  \end{table}

 Using the current integrated luminosity of the Belle experiment, we find that the discrimination power between different spin states exceeds 
 $5\sigma$ in all tested spin configurations. However, the discrimination power for distinguishing between different parity states with the same spin is relatively weaker. The rejection significances of $\frac{3}{2}^-$ over $\frac{3}{2}^+$ for the $\Xi(1530)^0$,  $\frac{1}{2}^+$ over $\frac{1}{2}^-$ for the $\Xi(1690)^0$, $\frac{3}{2}^+$ over $\frac{3}{2}^-$ for the $\Xi(1820)^0$, and $\frac{5}{2}^+$ over $\frac{5}{2}^-$ for the $\Xi(2030)^0$, are less than $3\sigma$, while the rejection significances of $\frac{1}{2}^+$ over $\frac{1}{2}^-$ for the $\Xi(1620)^0$ and $\frac{1}{2}^+$ over $\frac{1}{2}^-$ for the $\Xi(1950)^0$ are greater than $3\sigma$. This is because (1) the effective vertices for $\Xi_c^+ \to \Xi^{*0} \pi^+$ are identical for $P=\pm1$; and (2) the statistical sample size of the resonance affects the ability to discriminate between $P=\pm1$ states with the same spin. 

 To study the effect of statistical sample size on parity determination, we increased the toy MC sample statistics by a factor of five and repeated the analysis. The PWA results for the enlarged sample demonstrate that the rejection significance of the alternative parity hypothesis over the favored parity hypothesis exceeds $5\sigma$ for each tested resonance, indicating that the discrimination power improves with the increasing sample size. Under limited statistics, the primary challenge in parity determination arises from the identical effective vertices of the  $\Xi_c^+ \to \Xi^{*0} \pi^+$ decay for both $P=\pm1$ hypotheses. However, the effective vertices of the subsequent decay $\Xi^{*0}\to\Xi^- \pi^+$ exhibit distinguishable behavior for $P=\pm1$, enabling the model to perform parity determination effectively under large statistics. 

 The sources of systematic uncertainties include the background estimation, the detector resolution, and the choice of the form factor. We changed the background events by $\pm\sqrt{N_\text{bkg}}$ and repeated the PWA fit to estimate the uncertainty from the background, where $N_\text{bkg}$ is the number of background events. The uncertainty from the detector resolution was estimated by convolving a Gaussian function with a resolution of 2 MeV~\cite{confirm-xi1620}. We modified the form factor to the one used in Ref.~\cite{xi16901820-jp} to estimate the uncertainty from the different form factor choices. The favored spin-parities for each resonance from PWA fits under different situations are consistent with the input assignments. 

\section{Summary}
  This study demonstrates the capabilities and limitations of the current Belle dataset for analyzing the $\Xi_c^+ \to \Xi^-\pi^+\pi^+$ decay and determining the spin-parity of the $\Xi^{*0}$ resonances. Under the current integrated luminosity of the Belle experiment, we expect to see multiple $\Xi^{*0}$ states in the $M(\Xi^-\pi^+)$ mass spectrum, including the $\Xi(1530)^0$, $\Xi(1620)^0$, $\Xi(1690)^0$, $\Xi(1820)^0$, $\Xi(1950)^0$, and $\Xi(2030)^0$, based on the Belle publication~\cite{confirm-xi1620}. Among these, $\Xi(1620)^0$ and $\Xi(1950)^0$ are anticipated to exhibit relatively significant signal statistics, while $\Xi(1690)^0$, $\Xi(1820)^0$, and $\Xi(2030)^0$ are expected to have smaller signal statistics. The PWA method for the $\Xi_c^+ \to \Xi^-\pi^+\pi^+$ decay, implemented using the FDC package, is shown to be reliable based on the MC simulation. Measuring masses and widths, and establishing the favored $J^P$ hypotheses of these $\Xi^{*0}$ states are feasible with current statistics. The rejection significances for alternative spin hypotheses over favored spin hypotheses are all greater than $5\sigma$ for six $\Xi^{*0}$ states. Only $\Xi(1620)^0$ and $\Xi(1950)^0$ have the rejection significances for alternative parity hypothesis over favored parity hypothesis greater than $3\sigma$, owing to their larger signal statistics. A larger dataset, such as that from the Belle II experiment, which is expected to be approximately 5 times the integrated luminosity of the Belle, is required to enhance sensitivity to parity determination.

  Once the $J^P$ quantum numbers of these $\Xi^{*0}$ states are determined, physicists will deepen insights into their internal structures, especially for the $\Xi(1620)^0$ and $\Xi(1690)^0$. We strongly recommend the experimental physicists leverage the large datasets collected at Belle and Belle II experiments to perform the PWA of the $\Xi_c^+ \to \Xi^-\pi^+\pi^+$ process, with the aim of determining the $J^P$ values of the $\Xi^{*0}$ states in future studies. \\

%%%%%%%%%%%%%%%%%%%%%%%%%%%%%%%%%%%%%%%%%%%%%%%%%%%%%%%%%%%%%%%%%%%%%
%ACKNOWLEDGMENTS
\section*{ACKNOWLEDGMENTS}
We thank E.~Wang for reading the manuscript and for constructive comments and suggestions.
This work is supported in part by National Key Research and Development Program of China under Grant No. 2024YFA1610503, National Natural Science Foundation of China under Grant No. 12161141008, No. 12135005, and No. 12175244, China Postdoctoral Science Foundation (CPSF) under Grant No. 2024M760486, and China Postdoctoral Fellowship Program of CPSF under Grant No. GZC20230539. 

%%%%%%%%%%%%%%%%%%%%%%%%%%%%%%%%%%%%%%%%%%%%%%%%%%%%%%%
\section*{APPENDIX:ADDITIONAL DISTRIBUTIONS}
\begin{figure*}[htbp]
   \flushleft
    \includegraphics[width=0.98\textwidth]{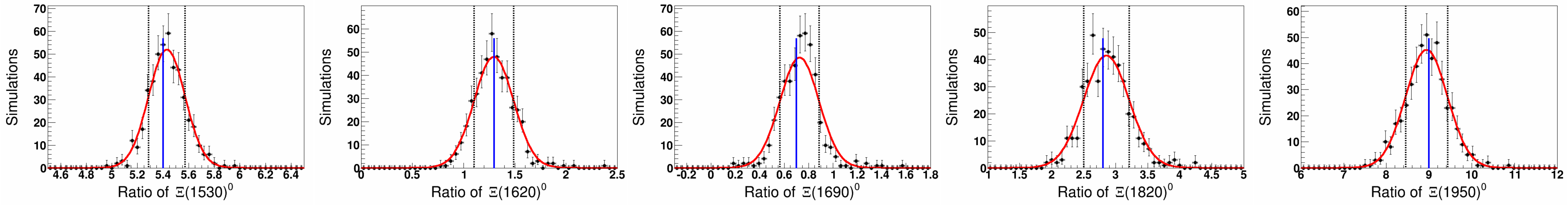}
    \includegraphics[width=0.98\textwidth]{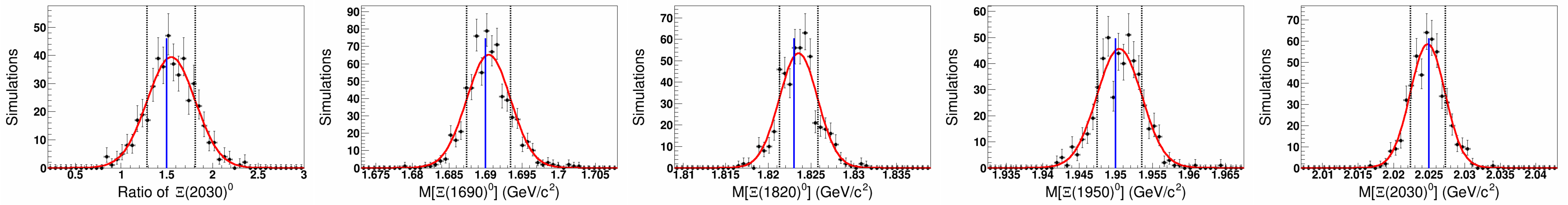}
    \includegraphics[width=0.98\textwidth]{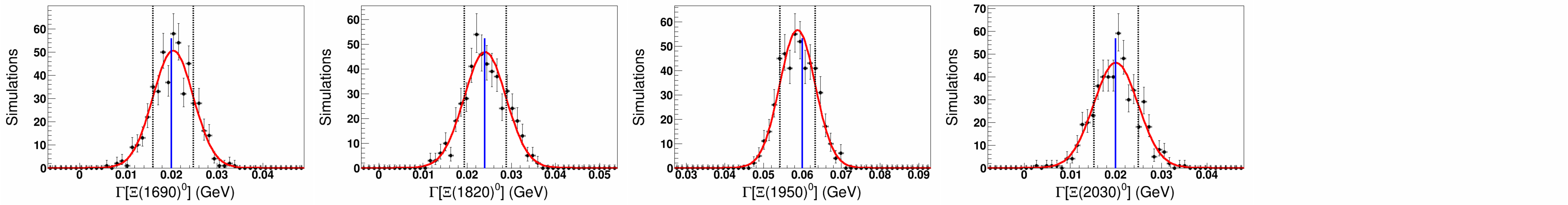}
    \caption{The distributions of the outputs of ratios, masses, and widths for the $\Xi(1530)^0$, $\Xi(1620)^0$, $\Xi(1690)^0$, $\Xi(1820)^0$, and $\Xi(2030)^0$. The dots with error bars are the output values, the red curves are the Gaussian functions, the two black dashed lines correspond to $\pm1$ standard deviation, and the blue lines represent the input values. }
    \label{fig:sec5:all-ioc}
    \end{figure*}

\begin{figure*}[htbp]
    \flushleft
    \includegraphics[width=0.98\textwidth]{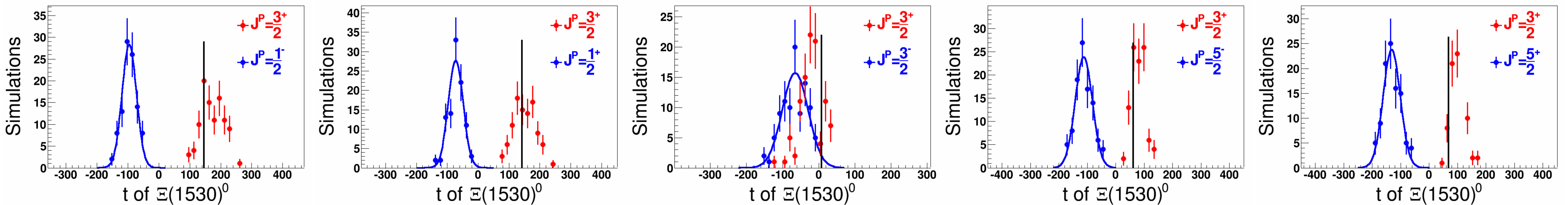}
    \includegraphics[width=0.98\textwidth]{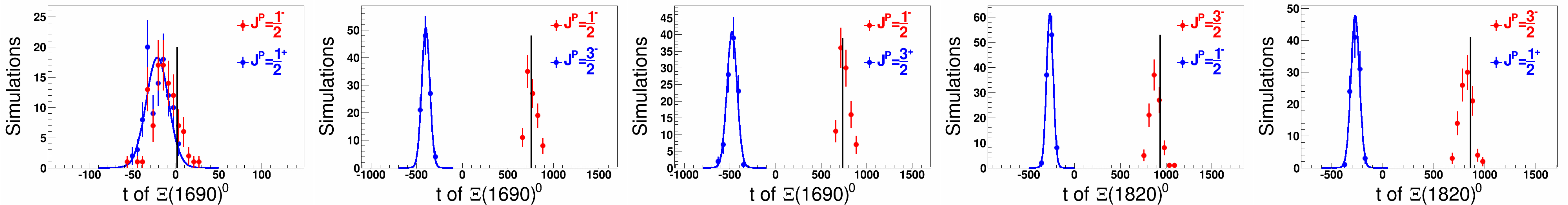}
    \includegraphics[width=0.98\textwidth]{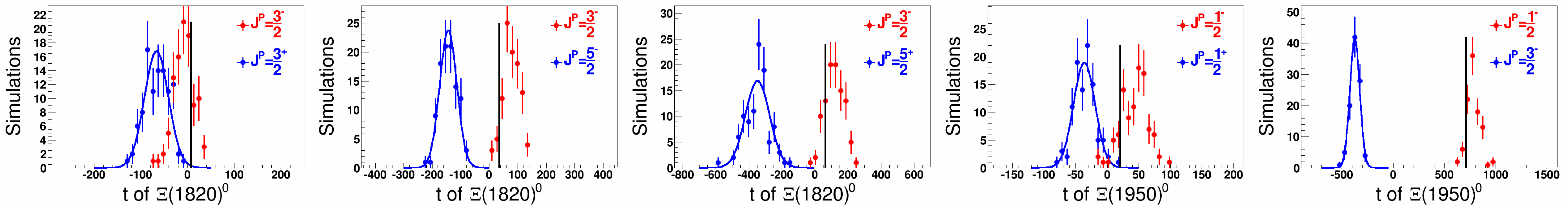}
    \includegraphics[width=0.98\textwidth]{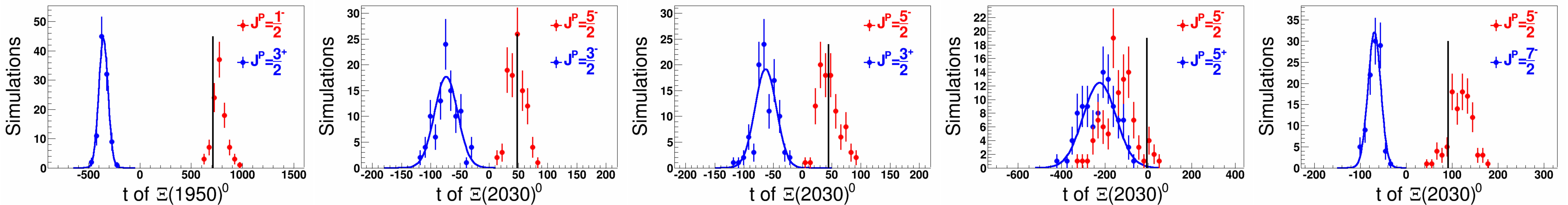}
    \includegraphics[width=0.98\textwidth]{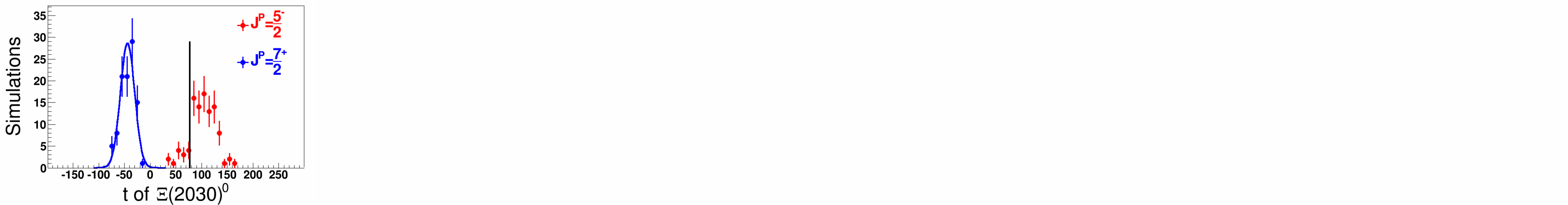}
    \caption{Distributions of the test statistic $t$ of the $\Xi(1530)^0$, $\Xi(1690)^0$, $\Xi(1820)^0$, and $\Xi(2030)^0$, for the simulated experiments under the $J^P_\text{alt}$ hypotheses (blue histograms) and under the $J^P_\text{fav}$ hypotheses (red histograms). The values of the test statistics for the toy data, $t_\text{toydata}$, are shown by the solid vertical lines. The curves are the Gaussian fitted distributions to the left peaks. }
    \label{fig:sec4:ttest-all}
  \end{figure*}

  Figure~\ref{fig:sec5:all-ioc} shows the distributions of the outputs of ratios, masses, and widths for the $\Xi(1530)^0$, $\Xi(1620)^0$, $\Xi(1690)^0$, $\Xi(1820)^0$, and $\Xi(2030)^0$. Figure~\ref{fig:sec4:ttest-all} shows the test statistic $t$ distributions for the $\Xi(1530)^0$, $\Xi(1690)^0$, $\Xi(1820)^0$, and $\Xi(2030)^0$. The simulations for the red dots with error bars are performed under the $J^P_\text{fav}$ hypothesis, while those in the blue dots with error bars correspond to the $J^P_\text{alt}$ hypothesis.

%%%%%%%%%%%%%%%%%%%%%%%%%%%%%%%%%%%%%%%%%%%%%%%%%%%%%%%
% Surround figure environment with turnpage environment for landscape
% figure
% \begin{figure}
% \end{figure}

% 
% \begin{table}%[H] add [H] placement to break table across pages
% \begin{ruledtabular}
% Lines of table here ending with \\
% \end{tabular}
% \end{ruledtabular}
% \end{table}

% Surround table environment with turnpage environment for landscape
% table
% \begin{table}
% \begin{ruledtabular}
% \end{tabular}
% \end{ruledtabular}
% \end{table}

%\appendix
%\section{}

%\begin{acknowledgments}
%\end{acknowledgments}


\end{document}